\documentclass[conference,usenatbib]{basi}
\usepackage[T1]{fontenc}
\usepackage[british]{babel}
\usepackage[varg]{txfonts}
\usepackage{rotating}
\usepackage{dcolumn}
\begin{document}
\title[Coherent Radio Emission from Pulsars]{Coherent Radio Emission from Pulsars}
\author[Mitra, Melikidze \& Gil]%
       {D. Mitra$^1$\thanks{email: \texttt{dmitra@ncra.tifr.res.in}},
        G. Melikidze$^{2,3}$,
        \& J. Gil$^2$,
       \\
       $^1$National Center for Radio Astrophysics, Pune, India\\
$^2$Kepler Institute of Astronomy, University of Zielona Gora, Lubuska 2, 65-265
Zielona G\'ora, Poland \\
$^3$Abastumani Astrophysical Observatory, Ilia State
University, 3-5 Cholokashvili Ave., Tbilisi, 0160, Georgia}
\pubyear{2015}
\volume{12}
\pagerange{\pageref{firstpage}--\pageref{lastpage}}

\date{Received --- ; accepted ---}

\maketitle
\label{firstpage}

\begin{abstract}
We review a physical model where the 
high brightness temperature of 10$^{25}-10^{30}$ K observed in
pulsar radio emission is explained by  
coherent curvature radiation excited in the relativistic electron-positron 
plasma in the pulsar magnetosphere.
\end{abstract}


\section{Introduction}\label{s:intro}
The pulsar magnetosphere
needs a minimum charge density (Goldreich
\& Julian 1968) of $n_{GJ} = \Omega . B/2\pi c$ (frequency $\Omega = 2\pi/P$, P is the pulsar period, $B$ 
is the magnetic field and $c$ is velocity of light) to maintain corotation.
The process by which plasma is generated and flows globally in the magnetosphere
is a matter of intense research (see Spitkovsky 2011), however 
in all models the pulsar
radio emission arises due to growth of plasma instabilities in the
relativistic electron positron plasma streaming along curved open dipolar magnetic field lines. 
Here we briefly discuss the basic observations that
constrain the pulsar radio emission, followed by a proposed model for
the radio emission.


\begin{figure}
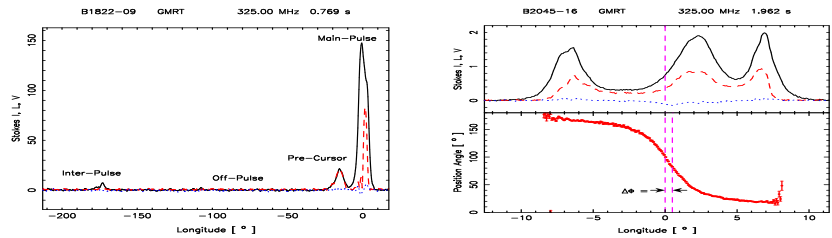

\centerline{\includegraphics[height=5cm,width=3cm,angle=-90]{mitra_fig1a.ps} \qquad
            \includegraphics[height=5cm,width=3cm,angle=-90]{mitra_fig1b.ps}}
\caption{The left panel shows the average pulse profile of PSR B1822-09 where the 
emission components like MP, IP, PC are seen. The red and blue
curves correspond to the linear and circular polarization. The right panel shows 
the MP for PSR B2045-16 (period 1.9 sec). Top panel: Total intensity (black), linear (red)
and circular (blue). Bottom panel shows the average polarization position angle (PPA). 
An S-shape as predicted by the RVM is visible.
The left dashed line runs through the center of the total intensity profile and the right dashed
line runs through the center (inflexion point) of the PPA traverse. The range $\Delta \phi \sim 0.5^{\circ}$ is due to A/R delay
which gives the radio emission height $h \sim \frac{c}{4} \frac{\Delta\phi P}{360^{\circ}}$ km as about 300 km (see \citet{bcw91,ml04}). \label{fig1}}

\end{figure}

\section{The radio observational Constraints and the model}\label{s:const}
The pulsar emission consists
of a bright main pulse (MP); sometimes an interpulse (IP) located
180$^{\circ}$ away from the MP; occasionally pre/post-cursor (PC)
emission connected via a bridge to the MP
and the unpulsed emission called off-pulse emission (see left Fig~\ref{fig1}).
The MP, IP and PC emission are highly polarized. The linear polarization
position angle (PPA) below the MP for a large number of pulsars are in very good
agreement with the rotating-vector model (RVM, Radhakrishnan \& Cooke
1969) which predicts that the PPA's are consistant with emission arising from open
dipolar field lines (see Fig.~\ref{fig1}). Here we will discuss the model that explains the
MP emission of normal pulsars having periods longer than 100 msec and
dipolar magnetic fields of $\sim10^{12}$ Gauss.
Two most crucial observations that constrain the MP emission are as follows.
(1) {\bf Emission heights:} The radio emission
originates at heights less than 10\% of the light cylinder implying
brightness temperatures of about $10^{25} - 10^{30}$ K.  An
example for measuring radio emission heights due to special
relativistic effect of aberration and retardation (A/R) is explained in
Fig.~\ref{fig1}.
(2) {\bf Polarization of outgoing radiation:} X-ray
and radio polarization provide
evidence that the emerging radiation is polarized parallel or
perpendicular to the dipolar magnetic field line planes (\citet{lcc01})

\begin{figure}
\centerline{
            \includegraphics[height=4cm,width=6cm]{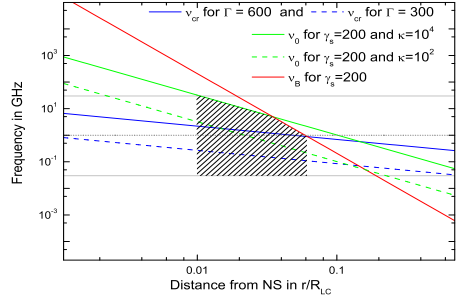}}
\caption{The figure shows the secondary plasma properties as a function of $\Re$ for a pulsar
with $P=1$ sec and $\dot{P}=10^{-15}$. Red line: cyclotron frequency, Green line: plasma frequency, Blue 
line: curvature radiation. The shaded region $\nu_{cr} < \nu_\circ$ is where radio emission can be generated (see Paper IV for details).
\label{fig2}}
\end{figure}

{\bf The Model:}Coherent curvature radiation has been considered as a natural emission
mechanism for the observed pulsar radiation (Ruderman \& Sutherland 1975, RS75). 
The basic physical idea in this model is the formation of an inner vacuum gap 
(IVG) near
the pulsar polar cap where non-stationary spark associated
relativistic ($\gamma_p \sim 10^6$) primary particles are
generated. These particles further radiate in strong magnetic field and
the photons thereby produce secondary e$^+$e$^-$ plasma with $\gamma_s
\sim 400$. Thus the charge density in the secondary plasma $n_s$ gets 
multiplied by a factor $\kappa = n_s/n_{GJ} \sim 10^{4}$. Growth of plasma instabilities in 
the secondary plasma leads to
formation of charged bunches (solitons) which can emit curvature
radiation due to acceleration in curved magnetic fields. However
several key issues in the RS75 model had remained unsolved for a long
time. For example observations of hot xray polar-caps and subpulse drift (not discussed here) 
has led to the conjecture that IVG is in fact partially screend gap (PSG model by Gil et al. 2004).
Further the theory of formation of charged bunches leading
to curvature radiation in plasma was not developed. Recently in a
series of paper Melikidze, Gil \& Pataraya (2000), Gil, Luybarsky \&
Melikidze (2004), Mitra, Gil \& Melikidze (2009) and Melikidze, Mitra
\& Gil (2014) (Paper I--IV hereafter) we have attempted to develop a self-consistant theory of
coherent curvature radiation for pulsars. Here we will
only discuss the condition in the secondary plasma that leads to 
coherent curvature radiation. 

Fig.~\ref{fig2} shows the plasma properties of secondary electron 
positron plasma as a fraction of distance $\Re$ to the light cylinder 
(i.e.: $\Re = h/R_{LC}$, where $R_{LC}= {Pc}/{2\pi}$) 
for a pulsar with $P$ of 1 second
and period derivative {$\dot{P}$ of 10$^{-15}$ s/s.
Note that the plasma frequency, cyclotron frequency and
the characteristic frequency of the soliton
curvature coherent radiation can be expressed in terms of $P$ and
{$\dot{P}$.
In the figure the plasma frequency $\nu_{\circ}$(in green), cyclotron frequency 
$\nu_{B}$ (in red) and
the characteristic frequency $\nu_{cr}$ of the soliton
curvature coherent radiation (in blue) is shown (see Paper I and IV). The shaded region
is the range where the pulsar radiation can be generated.

{\bf The mechanism of coherent radio emission}
The only plasma instability that can arise at altitudes lower than 10\% of
the light cylinder is the two-stream instability, 
as all other instabilities are suspended
by the strong magnetic field. The two-stream instability is a result
of the effective energy exchange between particles and waves, that can
occur if the phase velocity of waves ($\omega/k$) is near to the
velocity of resonant particles ($v_r$), i.e. the resonant condition
$(\omega - kv_r)=0$ is satisfied. 
Such conditions can naturally be realized if a plasma is
produced via non-stationary gap discharge. 
The spark discharge timescale in the PSG is a few tens of microseconds 
and this results in
overlapping of successive clouds of outflowing secondary plasma. Each
elementary spark-associated plasma cloud has a spread in momentum and
the overlapping of particles with different momentum leads to two
stream instability in the secondary plasma cloud.
This triggers strong Langmuir
turbulence in the plasma and if this turbulence is strong enough, the
waves become modulationally unstable. The unstable wave packet
described by the nonlinear Schr\"{o}dinger equation leads to formation
of a quasi-stable nonlinear solitary wave, i.e. a soliton (Paper I). 
The longitudinal (along ${\bf k}$) size of the soliton
should be much larger than the wavelength of the linear Langmuir wave.
Also it has to be charged to be able to radiate coherent curvature
emission. Thus, the soliton bunch has to be charge separated, which
can be caused either by difference in the distribution function of
electrons and positrons, or by admixture of iron ions in the secondary
plasma or by both these effects. A sufficient number of charged
solitons is formed which can account for the observed radio luminosity
in pulsars (Paper I). The wavelength of the emitted waves should be
longer than the longitudinal size of the soliton $\Delta$. This is the
necessary condition for the coherency of a curvature radiation
process. Thus, the frequencies plotted in Fig.~\ref{fig2} should
obey the following constraints $\nu_{\rm cr}<\frac{c}{\Delta}\ll
\nu_{\circ} \ll \nu_{B}.$ It is clearly seen from
Fig.~\ref{fig2} that the observed pulsar radiation cannot be
generated at altitudes exceeding 10\% of the light cylinder radius
(practically the radio emission region should be contained between one
to several percent of $R_{LC}$; see dashed area in Fig.~
\ref{fig2}). This conclusion is based purely on the properties of
plasma and the emission mechanism. And it corresponds perfectly to the
other limits on emission heights obtained from observations. 
\citep[e.g.][]{bcw91,ml04}.

The curvature radiation excites the extraordinary X-mode (polarized $\perp$ to 
the magnetic field plane) and the ordinary 
O-mode (polarized $\parallel$ to the magnetic field plane) in the secondary plasma, and recently it was shown in Paper IV that the
radiation can emerge retaining its polarization without getting affected 
due to propagation effects in the plasma (Paper IV). This X and O mode should hence
be associated with the parallel and perpendicularly polarized emission observed
in pulsars.   
\section*{Acknowledgements}
DM thanks the RETCO organizers for invitation to give
a talk in the meeting.


\label{lastpage}

\begin{thebibliography}{}
\bibitem[Blaskiewicz, Cordes \& Wasserman(1991)]{bcw91} Blaskiewicz, M., Cordes, J. M., \& Wasserman, I., 1991, ApJ, 370, 643
\bibitem[Gil, Melikidze \& Geppert(2003)]{gmg03} Gil, J., Melikidze, G. I. \& Geppert, U., 2003, A\&A, 407, 315
\bibitem[Gil, Lyubarsky \& Melikidze(2004)]{glm04} Gil, J., Lyubarsky, Y., \& Melikidze, G. I., 2004, ApJ, 600,872 (Paper II)
\bibitem[Golredich \& Julian (1968)]{gj68} Goldreich, P \& Julian W. H.,1968, ApJ, 157, 869 
\bibitem[Lai, Chernoff \& Cordes (2001)]{lcc01} Lai, D., Chernoff, D. F., \& Cordes, J. M., 2001, ApJ, 549, 1111
\bibitem[Melikidze, Gil \& Pataraya(2000)]{mgp00} Melikidze, G. I, Gil, J., \& Pataraya, A. D. 2000, ApJ, 544, 1081 (Paper I)
\bibitem[Melikidze, Mitra \& Gil (2014)]{mmg14} Melikidze, G. I., Mitra, D. \& Gil, J., 2014, 794, 105 (Paper IV)
\bibitem[Mitra, Gil \& Melikidze(2009)]{mgm09} Mitra, D., Gil, J. \& Melikidze, G. I., 2009, ApJ, 696L, 141 (Paper III)
\bibitem[Mitra \& Li(2004)]{ml04} Mitra, D. \& Li, X. H., 2004, A\&A, 421, 215
\bibitem[Radhakrishnan \& Cooke(1969)]{rc69} Radhakrishnan V., \& Cooke D. J., 1969, ApJL, 3, 225
\bibitem[Ruderman \& Sutherland(1975), RS75 hereafter]{rs75} Ruderman, M. A., \& Sutherland, P. G., 1975, ApJ, 196, 51
\bibitem[Spitkovsky (2011)]{s11}Spitkovsky, A, 2011, HEEP, editor Torres, D. F \& Rea, N., 139
\end{thebibliography}
\end{document}